\documentclass[12pt,preprint,notoc,nohyper]{AKRSX} 

\usepackage{epsfig,multicol,bbm}

\newcommand\fverb{\setbox\pippobox=\hbox\bgroup\verb}
\newcommand\fverbdo{\egroup\medskip\noindent%
            \fbox{\unhbox\pippobox}\ }
\newcommand\fverbit{\egroup\item[\fbox{\unhbox\pippobox}]}
\newbox\pippobox

\title{Classification of spacetimes according to conformal Killing vectors}

\author{K. Saifullah\thanks{\emph{On leave from:}
Centre for Advanced Mathematics and Physics, National University of
Sciences and Technology, Rawalpindi, Pakistan, \emph{and} Department
of Mathematics, Quaid-i-Azam University, Islamabad, Pakistan.} \\
    School of Mathematical Sciences, Queen Mary, University of London, \\ London,
UK\\
    Electronic address: \email{saifullah@qau.edu.pk}}

\preprint{}  

\abstract{Conformal Killing vectors (CKVs) preserve the spacetime
metric up to a factor. Homothetic vectors and Killing vectors are
special cases of CKVs. Classification of some classes of spacetimes
on the basis of CKVs give interesting results showing how homothetic
and Killing vectors which form subsets of the set of CKVs can be
recovered as a result of the above classification.}

\dedicated{}

\begin{document}

Einstein's theory of general relativity is based on the realization
that geometry, represented by the Riemann curvature tensor
$R_{bcd}^{a}$ of the spacetime can be related to the distribution
and motion of matter, denoted by the stress-energy tensor $T_{ab}$.
This relation is explained by Einstein's field equations (EFEs),

\begin{equation}
\ R_{ab}-\frac{1}{2}Rg_{ab}=\kappa T_{ab} \ \ \ \ \ \ \left(
a,b=0,1,2,3\right) . \label{s1}
\end{equation}
Here $g_{ab}$ is the metric tensor, $R_{ab}$ the Ricci tensor, $R$
the Ricci scalar and $\kappa =\frac{8\pi G}{c^{2}}$, where $G$ and
$c$ are the gravitational constant and the speed of light
respectively. (We have ignored the term with the cosmological
constant.) Metric, $g_{ab}$, is the dynamical quantity in EFEs which
varies over the spacetime. EFEs (\ref{s1}) break down into ten
highly non-linear differential equations and so far very few exact
solutions have been discovered by imposing certain restrictions
\cite{2r}. One of such restrictions could be to allow a spacetime to
admit certain symmetry properties. For example, the isometry group
$G_{m}$ of $\left(M, \mathbf{g}\right)$ is the Lie group of smooth
maps of manifold $M$ onto itself leaving $\mathbf{g}$ invariant. The
subscript $``m"$ is equal to the number of generators or isometries
of the group. It is the Lie algebra of continuously differentiable
transformations $K^{a}\partial /\partial x^{a}$ where $K^{a}=$
$K^{a}\left( x^{b}\right) $ are the components of the vector field
$\mathbf{K}$ known as a Killing vector (KV) field. In other words, a
KV field $\mathbf{K}$ is a field along which the Lie derivative of
the metric tensor $\mathbf{g}$ is zero i.e. $\pounds
_{\mathbf{K}}\left( g_{ab}\right)=0.$

In addition to isometries there are other types of motions which are
even more restrictive and therefore could be more useful as far as
the solution of Eqs.(\ref{s1}) and their properties are concerned.
For example, the study of homothetic vectors (HVs) and conformal
Killing vectors (CKVs) are significant in general relativity
\cite{9r}. CKVs are motions along which the metric tensor of a
spacetime remains invariant up to a scale i.e.

\begin{equation}
\pounds _{\mathbf{\xi }}g_{ab}=g_{ab,_{c}}\xi ^{c}+g_{ac}\xi
_{,b}^{c}+g_{bc}\xi _{,a}^{c}=2\phi g_{ab} \ .  \label{s2}
\end{equation}

Conformal motions are determined by the arbitrary constants
appearing in the vector field $\mathbf{\xi }=\xi ^{a}\partial /
\partial x^{a}$ when $\phi =$ $\phi \left( t,x,y,z\right)$. In
the above equation, \textquotedblleft $,$ \textquotedblright
represents derivative with respect to coordinates $x^{a}$. If $\phi$
is constant $\mathbf{\xi}$ represents HVs and if it is zero, we
simply get the KVs. It is clear from the definition that HVs and KVs
are special cases of CKVs. The study of the symmetry groups of a
spacetime is a useful tool not only in constructing spacetime
solutions of EFEs but also for classifying the known solutions
according to the Lie algebras, or structure generated by these
symmetries. Previously, CKVs have been studied for various
spacetimes like Minkowski \cite {5r}, Robertson-Walker \cite{6r} and
pp-waves \cite{7r}.

Important results regarding the dimensionality of these symmetries
include (see, for example,  Refs. 2, 6):

\textbf{1.} Riemannian space $V_{n}$ admits a group of motions
$G_{m}$ where $m\leq n\left( n+1\right)/2$.

\textbf{2.} A Riemannian space $V_{n}$ cannot admit a maximal group
of motions $G_{m}$ where $m=n\left( n+1\right)/2-1.$ If a spacetime
admits a $G_{m}$ as the maximal group of isometries then the HVs
group $H_{r}$ is at the most of order $r=m+1$.

\textbf{3.} The set of conformal vector fields on $M$ is
finite-dimensional and its dimension is less then or equal to $15$.
If this maximum number is attained, the spacetime is conformally
flat. If it is not conformally flat then the maximal dimension is
$7$.

Let us consider, for example, the class of spherically symmetric
spacetimes which, in the usual coordinates, with $\nu \left(
t,r\right)$, $\lambda \left( t,r\right)$ and $\mu \left( t,r\right)$
as arbitrary functions, can be written as

\begin{equation}
ds^{2}=-e^{\nu \left( t,r\right) }dt^{2}+e^{\lambda \left(
t,r\right) }dr^{2}+e^{\mu \left( t,r\right)}(d\theta ^{2}+\sin
^{2}\theta d\varphi ^{2}) \ . \label{a1}
\end{equation}
These spacetimes admit 3 KVs

\begin{eqnarray}
K^{1} &=&\sin \phi \frac{\partial }{\partial \theta }+\cos \phi \cot
\theta \frac{\partial }{\partial \phi } \ ,  \nonumber \\
K^{2} &=&\cos \phi \frac{\partial }{\partial \theta }-\sin \phi \cot
\theta \frac{\partial }{\partial \phi} \ ,   \nonumber  \\
K^{3} &=&\frac{\partial }{\partial \phi} \ . \nonumber
\end{eqnarray}
In the static case these admit a timelike KV, $K^{4}=\partial
/\partial t$, also. The classification of HVs of spherically
symmetric spacetimes admitting maximal isometry groups larger than
$SO\left( 3\right)$ was obtained along with their metrics \cite{12r}
by using the homothety equations and without imposing any
restriction on the stress-energy tensor. The possible maximal
homothety groups $H_{r} $ for these spacetimes are of the order
$r=4,5,7,11$; for $r=11$, the only spacetime is Minkowski. The
general solution and classification of conformal motions for these
spacetimes \cite{11r} shows that the group of CKVs is $G_{4+n}$
where $n$, the number of CKVs, is either $2$ or $11$. In the case
$n=2$, both CKVs are necessarily proper. For the conformally flat
case, up to $6$ of the $11$ CKVs may be improper.

For the plane symmetric metric

\begin{equation}
ds^{2}=-e^{\nu \left(t,x \right) }dt^{2}+e^{\lambda \left(t, x
\right) }dx^{2}+e^{\mu \left(t,x \right) } \left(dy^2+dz^2\right)
 \ , \label{planemetric}
\end{equation}
the minimal symmetry is given by

\begin{equation}
K^{1}=\frac{\partial}{\partial y} \ , \
K^{2}=\frac{\partial}{\partial z} \ , \
K^{3}=z\frac{\partial}{\partial y}-y\frac{\partial}{\partial z} \ .
\nonumber
\end{equation}
In the static case the spacetimes admit a timelike KV,
$K^{4}=\partial /\partial t$, in addition to the KVs given above.
The orders of the isometry groups for the associated metrics are 4,
5, 6, 7 and 10; 8 and 9 are not admissible \cite{plnkv}. Hence the
possible groups for HVs \cite{4r} are of the order 5, 6, 7 or 11.
Classification of these spacetimes according to CKVs \cite{shair} is
also in accordance with the established results.

\acknowledgments

The author is grateful to George Alekseev for helpful comments. A
research grant from the Higher Education Commission of Pakistan is
gratefully acknowledged. The author is also thankful to the National
University of Sciences and Technology, Pakistan for the travel
support to deliver this talk at MG11, Berlin, 2006.

\end{document}